\documentstyle[epsf,amstex,epsfig]{mn}
\loadboldmathitalic 
\title []{Revisiting the Cosmic Cooling Crisis}

\author[Balogh \etal]
{Michael L. Balogh$^{1,3}$, Frazer. R. Pearce$^{1}$, Richard G. Bower$^{1}$ \& Scott T. Kay$^{2}$\\
$^{1}$Department of Physics, University of Durham, South Road, Durham, DH1 3LE, UK\\
$^{2}$Astronomy Centre, CPES, University of Sussex, Falmer, Brighton, BN1 9QJ, UK\\
$^{3}$email:M.L.Balogh@@durham.ac.uk\\
}
\date{\today}
\def\lesssim{\mathrel{\hbox{\rlap{\hbox{\lower4pt\hbox{$\sim$}}}\hbox{$<$}}}}
\def\gtrsim{\mathrel{\hbox{\rlap{\hbox{\lower4pt\hbox{$\sim$}}}\hbox{$>$}}}}
\def\ion#1#2{#1$\;${\small\rm\@roman{#2}}\relax}
\def\etal{{\it et al.\thinspace}}
\def\Msun{\hbox{$\rm\thinspace M_{\odot}$}}
\def\fgcool{$f_{\rm c,global}$}
\def\fccool{$f_{\rm c,cluster}$}
\def\fcool{$f_{\rm c}$}

\begin{document} 
\maketitle 
\begin{abstract}
Recent measurements of the
$K-$band luminosity function now provide us with strong, reliable
constraints on the fraction of baryons which have cooled.  
Globally, this fraction is only about 5\%, and there is no strong
evidence that it is significantly higher in clusters.
Without an effective sub-grid feedback prescription, the cooled gas fraction
in any numerical simulation exceeds these observational constraints,
and increases with increasing resolution.
This compromises any discussion of galaxy and cluster properties
based on results of simulations which include cooling but do not
implement an effective feedback mechanism.
\end{abstract}

\section{Introduction}\large
Gas cooling in the expanding Universe is an intrinsically unstable
process because cooling acts to increase the density of the gas, which
in turn increases the cooling rate. The consequence of this is that as
soon as the gas within a system is able to cool at all it tends to
do so catastrophically, only regulated by the speed at which the gas
can respond to the new configuration.  Systems which collapse at low
redshift have a low mean density and, thus, long cooling times which
generally exceed their dynamical times.  Systems collapsing at higher redshifts have a
higher mean density and, because the cooling rate depends strongly upon
the gas density, they cool much more rapidly. 
In a Universe dominated by cold dark matter (CDM), the hierarchical
growth of structure consequently results in almost all of the baryons cooling
by the present day, unless additional physics is considered \cite{Cole91,WF91,BVM}.

Unfortunately, the naive theoretical prediction that all the baryonic
material in the Universe should have cooled into galaxies and formed
stars conflicts with the observation of X-ray emission from galaxy
clusters. This emission demonstrates that large amounts of hot gas
persist in the Universe. This material can arise from two possible
sources: either it was never contained within a collapsed halo in which
cooling was efficient, or energy injection (for instance feedback of
energy due to supernovae) has either reheated or prevented the
cooling of the gas.
The commonly used theory of Press \& Schecter \shortcite{PS} (see also
Bower 1991\nocite{PSext})  implicitly
assumes that all the matter in the Universe is contained within
haloes. The best numerical models \cite{Jenkins} indicate
that the fraction of material contained within haloes 
is certainly high, and only leave room
for a small uncollapsed fraction. 
Therefore, we appeal to feedback mechanisms \cite{Larson74,WR78,Cole91,WF91}
to reduce the amount of cold gas within collapsed haloes.


In this paper we review the observational constraints on the density
of cold and hot baryons, both globally and in rich clusters
(\S\ref{sec-obs}).  These data establish the global
cooled gas fraction at \fgcool$\approx0.073h$, and there is no
convincing evidence that it is much higher in clusters. We then 
examine cosmological simulation results 
in \S\ref{sec-sims}, and show that they can
easily cool far too much gas to be consistent with these data.
Finally, in \S\ref{sec-disc} we discuss the necessity of combining
simulations with analytic, sub-grid physics models, and the merits of
various initial attempts to do this.

Throughout this paper, we parametrise the Hubble constant as $H_\circ=100h$ km s$^{-1}$ Mpc$^{-1}$.  
We will need to consider two cosmologies, $\Lambda$ CDM ($\Omega_\circ=0.3$, $\Lambda=0.7$)
and standard CDM ($\Omega_\circ=1$, $\Lambda=0$).
Matter densities are given relative to the critical density, and mass-to-light ratios
($M/L$) are in solar units.

\section{Observational Constraints}\label{sec-obs}
\subsection{The Global Fraction of Cold Baryons}\label{sec-global}
The mass density of cooled baryons in the form of stars, $\Omega_{\rm stars}$,
can be obtained from the observed luminosity density 
(e.g. Blanchard, Valls-Gabaud \& Mamon 1992\nocite{BVM}) as long as 
1) we do not
miss a significant contribution from low surface brightness galaxies;
and 2) the mass-to-light ratio is well understood.  
Fortunately, surface brightness is tightly
correlated with total luminosity, and there does not appear to be a
significant contribution to the baryon density from low surface
brightness galaxies (Driver 1999\nocite{Driver}; Cole \etal 2000\nocite{Cole-2mass}).
Although there is an uncomfortably large range
in the Schechter parameters of the local optical luminosity function (due to
the large degeneracy between $\phi^*$ and $L^*$), Fukugita,
Hogan \& Peebles (1998, hereafter FHP)\nocite{FHP} demonstrate that
the total integrated $B-$band luminosity is constrained to within about 15\%.
However, the use of the optical 
luminosity function requires large and uncertain corrections for the
$M/L_B$ ratio, which is sensitive to stellar populations and, thus, must
be determined separately for different types of galaxy. 

A more robust estimate is obtained from the $K-$band luminosity function, since $M/L_K$
is less sensitive to star formation history, varying by only a factor of about
two due to stellar population age differences \cite{BdJ}.  
Recently, Cole \etal \shortcite{Cole-2mass} have computed the mass function
using over 17000 galaxies with redshifts from the 2dF galaxy redshift survey
and magnitudes from the 2-micron all sky survey (2MASS).  
From the galaxy colours and stellar population synthesis models 
they determine $M/L_K$ for each galaxy (independent of $h$).  Their ($\Lambda CDM$) result is most sensitive
to the assumed initial mass function (IMF):  for a Kennicutt \shortcite{K83}
IMF, $\Omega_{\rm stars}=0.0014 h^{-1}$, while for the Salpeter \shortcite{Sp}
IMF, $\Omega_{\rm stars}=0.0026 h^{-1}$.

The average stellar $M/L_K$ from which Cole \etal\ \shortcite{Cole-2mass} derive $\Omega_{\rm stars}$
is 0.73 and 1.32 for the Kennicutt and Salpeter IMFs, respectively (independent
of $h$).
A substantially higher $M/L_K$ would require a large contribution from low mass stars
and brown dwarfs.  Although these objects are likely to be a significant component of the mass
budget, there is no strong evidence that they are more abundant than
expected from standard IMFs \cite{FJF,Gizis,LR}.  
In particular, Reid \etal \shortcite{Reid} claim that brown dwarfs
contribute no more than 15\% of the mass of the Galactic disk.
Earlier microlensing results which suggested that there might be a
large population of dark objects in the Galaxy halo are no longer 
compelling, following a reanalysis of the MACHO project data (Alcock
\etal 2000)\nocite{MACHO}.  

Dynamical measurements
of early type galaxies \cite{vdM,BdJ} favour $M/L_K\lesssim1.1h$; this is an
upper limit to the global average because later type galaxies have lower $M/L_K$.  
We therefore prefer the estimate of $\Omega_{\rm stars}=0.0014 h^{-1}$
based on the Kennicutt \shortcite{K83} IMF with a mean $M/L_K=0.73$.

Cooled gas is also present in the form of neutral and molecular gas,
but these make only small contributions.  The density of neutral gas is
$\Omega_{\rm atomic}\approx0.000188 h^{-1}$, as recently calculated by
the HI Parkes All Sky Survey (HIPASS) team \cite{HIPASS}.
For the amount of gas in molecular form, we adopt the relation
$\rho_{H_2}/\rho_{HI}=0.81$ used by FHP, gleaned from the CO-based
compilation of Young \& Scoville \shortcite{YS}.  Together, the
neutral and molecular gas only represent $\sim 10$\% of the stellar
mass, which is the fraction we will adopt, so that $\Omega_{\rm cold}=1.1\Omega_{\rm stars}$.


The total baryon content of the universe is unknown observationally, because baryons
in the warm plasma phase believed to occupy normal galaxy haloes are very difficult
to see \cite{B+99}.  However, if nucleosynthesis calculations are correct,
the baryon density $\Omega_b$ can be determined from deuterium abundances.
The best current estimate, $\Omega_b=0.019h^{-2}$ \cite{BBN,BBN2}, therefore implies $f_{\rm c,global}=\Omega_{\rm
cold}/\Omega_b=0.073h$.   There are indications, from recent analysis of the 
combined BOOMERANG and
MAXIMA data, that $\Omega_b$ is much larger than this, $\Omega_b=0.039h^{-2}$ (Jaffe \etal 2000)\nocite{B+M},
which would reduce \fgcool\ by a factor of two.  

In summary, only about 5\% of the available baryons in the universe have cooled
(for $h\approx0.7$ and a Kennicutt IMF).  This result is slightly lower than the
minimum of the range
$0.062<f_{\rm c,global}<0.167$ found by FHP from optical data, but much more certain as it does not
rely on knowing the relative abundances of different galaxies (disks, spheroids
and irregulars) with very different (and uncertain) $M/L_B$ ratios.
As we will show in \S\ref{sec-sims}, this low value is in stark contrast
to the results obtained in numerical simulations which do not employ
an adequate feedback model.
%

\subsection{The Abundance of Hot Baryons}\label{sec-hidden}
In the previous section we claimed that most of the baryons in the Universe are
in a warm or hot phase, which is difficult to observe because 
it is not hot enough to be observable in
X-ray radiation \cite{BVM,Dave,B+99}.  
The existence of such a warm component is compatible with constraints on the
anisotropy of the microwave background, a long as the gas temperature satisfies
the constraint $T<4h \times 10^7$ K (at $z<1$) \cite{Wright}.

In clusters, however, the same gas
is much hotter, and is directly observable.  
As originally shown by White \etal \shortcite{WNF}, the total baryon content
of rich clusters, including this plasma, is fully consistent with $\Omega_b$ from element abundance
determinations, if $\Omega_\circ\approx0.3$.
FHP compile data on the baryon contributions in clusters due
to stars, cold gas and plasma, and demonstrate that
$\Omega_b/\Omega_\circ=0.112\pm0.05$ ($h=0.7$).  In this case, $\Omega_b$ is a sum
of all {\it observed} baryons in clusters.  Hence, for
$\Omega_\circ=0.3\pm0.1$, we have that $\Omega_b=0.034 \pm 0.019$ ($h=0.7$) which is
consistent with the baryon fraction deduced from deuterium abundances
\cite{BBN}.   Thus, the fact that $\Omega_{\rm stars}\ll \Omega_b$ in the Universe is almost
certainly a reflection of the fact that the warm gas has not been directly observed.

\subsection{The Dependence on Halo Mass}\label{sec-clusters}
Although we have demonstrated that \fgcool$\approx0.073h$, it is possible
that the cooled fraction in some environments, rich clusters for example, could be
very different.  In particular, Bryan \shortcite{Bryan} claims that the efficiency of
galaxy formation is dependent on halo mass, with more gas cooling in small haloes,
relative to large ones.
Most of the cluster data in the
literature has been obtained at optical wavelengths, so the correction for 
$M/L$ variations is not as robust as for the global value computed from the 
$K$-band field luminosity function in \S\ref{sec-global}. 
We will use the optimal values from FHP, who compile data from various sources,
including dynamical measurements and population synthesis model estimates.
The $M/L_B$ depends strongly on morphological type, and FHP find $M/L_B=6.5^{+1.8}_{-2.0}$ for E/S0 galaxies;
$M/L_B=1.5\pm0.4$ for spiral galaxies; and $M/L_B=1.1\pm0.25$ for irregular galaxies.
For the morphological composition of a ``typical'' cluster, they determine
$M/L_B=4.5 \pm 1$, and we will use this number.
Assuming a pure elliptical population
would increase stellar mass estimates by $\sim 45$\%. 
 

For low mass clusters and groups, the X-ray data only extend out to a small
fraction of the virial radius.  This can lead to a strong bias, since the gas distribution
is less concentrated than the stars \cite{RSB}.  We will consider the cluster and group data 
recently compiled by Roussell \etal \shortcite{RSB} specifically for the purpose
of addressing this issue.
In that paper, care was taken to ensure that stellar
and gas masses are computed to the same radius, and only those clusters for which
the X-ray data extend to at least 25\% of the virial radius are considered here.
Nonetheless, we warn that the data for groups with $kT<5$ keV 
must be treated with caution, as their gas masses still require a
substantial extrapolation to the virial radius.
In addition to the standard analysis of the X-ray data, Roussel \etal\ also 
consider hot gas mass estimates based on simulation-motivated scaling laws, which
do not require an assumption of isothermal, $\beta-$model profiles.
We use the measurements made based on the scaling laws of Bryan \& Norman \shortcite{BN}.
In addition, we consider the 15 rich clusters with reliable velocity dispersions
from the CNOC sample  \cite{CNOC1,CYE}, with X-ray data taken from Lewis \etal\ \shortcite{L+99}.

The stellar masses are computed from the integrated stellar luminosity 
function, assuming $M/L_B=4.5$, and corrected for undetected galaxies by extrapolating
the luminosity function with an assumed faint end slope of $\alpha\approx -1.2$.  
Finally, we
assume that the mass of cold baryons is 10\% larger than the stellar
mass, to account for the presence of neutral and molecular gas (see \S~\ref{sec-global}).

\begin{figure}
\begin{center}
\leavevmode \epsfysize=8cm \epsfbox{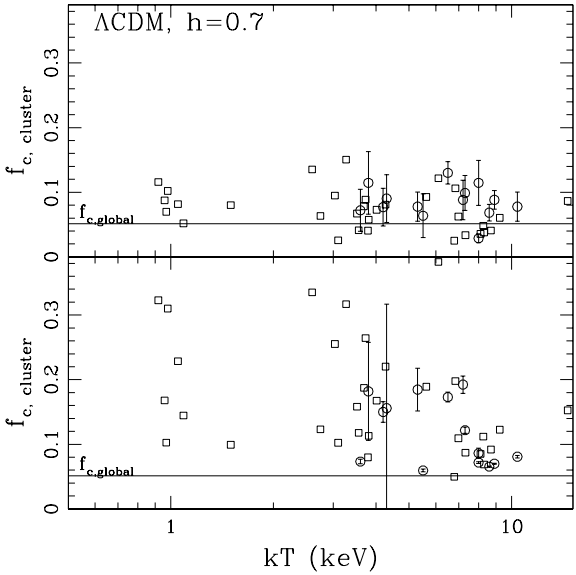}
\end{center}
\caption{The fraction of baryons in the cold phase in clusters, as a
function of X-ray temperature.  The data are from 
Carlberg \etal (1996, circles)
and Roussel \etal (2000, squares).  1$\sigma$ error bars are only available for the Carlberg \etal\ 
sample. The cluster data are
all normalized to a stellar $M/L_B=4.5$; increasing this number increases \fccool\ 
proportionally.    
The horizontal line shows the best estimate of the global cold
baryon fraction, \fgcool, based on the $K-$band luminosity function results of Cole \etal (2000);
note that \fgcool$\propto h$.  
{\it Bottom panel: }\fccool\ is computed from Equation \ref{eqn-fcool1},
where the total baryon mass is the sum of observed galaxies and hot gas.
The data points scale approximately like \fccool\ $\propto h^{0.5}$.
{\it Top panel:}\fccool\ is computed from Equation \ref{eqn-fcool2},
which does not depend on a measurement of $M_{\rm hot}$, but is
dependent on $\Omega_\circ$.  In this case, \fccool\ $\propto \Omega_\circ h$.
\label{fig-clus_obs}}
\end{figure}

We consider two approaches to compute \fccool\ from $M_{\rm cold}$ in the
clusters.  In the first case, we compute
\begin{equation}\label{eqn-fcool1}
f_{\rm c,cluster}=M_{\rm cold}/(M_{\rm cold}+M_{\rm hot}),
\end{equation}
where $M_{\rm hot}$ is determined from the observable X-ray emission from
the hot plasma, extrapolated to the virial radius. 
This will be valid if
the extrapolation is accurate (less likely for lower mass clusters), 
and will be an overestimate of \fccool\ if
there is an unobserved, warm component which does not contribute to the
X-ray emission (as is the case in galactic mass haloes).  We show \fccool\ for the
two samples in the bottom panel of Figure \ref{fig-clus_obs}, for a $\Lambda CDM$ cosmology with
$h=0.7$.
Both $M_{\rm cold}$ and $M_{\rm hot}$ are distance dependent measurements, with 
$M_{\rm cold} \propto h^{-1}$ and $M_{\rm hot} \propto h^{-1.5}$.
Since, usually, $M_{\rm hot}\gg M_{\rm cold}$, $f_{\rm cool} \propto h^{0.5}$ when calculated from Equation \ref{eqn-fcool1}.

Most of the \fccool\ values in the bottom panel of Figure \ref{fig-clus_obs} are higher than the global
fraction \fgcool\ computed in \S\ref{sec-global}, shown
as the horizontal line.  There is only weak evidence for a trend, in the sense that
\fgcool\ is lower in higher temperature systems.  However, this trend is much weaker than
suggested by Bryan \shortcite{Bryan} and, we caution that the $kT<5$ keV groups may still
be biased toward high fractions because of the limited extent of the X-ray data; none of these
groups have X-ray emission detected beyond half the virial radius.
Also, both the random and systematic uncertainties in observationally determining 
$M_{\rm hot}$, especially for the low temperature groups, are large.
This is emphasized in the top panel of Figure \ref{fig-clus_obs}, in which the same data are plotted,
but \fccool\ is computed as 
\begin{equation}\label{eqn-fcool2}
f_{\rm c,cluster}={M_{\rm cold} \over M_{\rm total}}{\Omega_\circ \over \Omega_b},
\end{equation}
where $M_{\rm total}$ is the total dynamical mass of the cluster,
and we take $\Omega_b$ from Burles \& Tytler \shortcite{BBN}.  This does not depend on the 
difficult measurement of $M_{\rm hot}$, but does require an estimate of $\Omega_\circ$, which is
0.3 in our $\Lambda CDM$ cosmology.  Note that, in this case, \fccool$\propto \Omega_\circ h$.
Not only has the apparent trend in \fccool\ with temperature now virtually
disappeared, but the scatter has been reduced considerably.  In this case the \fccool\ values
are in much closer agreement with \fgcool; the remaining difference is not
very compelling, since a slightly lower $\Omega_\circ\approx0.2$ would result
in concordance between the cluster and global fractions.  


These results differ from those of Bryan \shortcite{Bryan}, 
who claims that \fccool\ increases from cluster to group scales.
However, this trend could well be the result of a bias due to the limited
radial extent of the X-ray observations, $R_x$,  in the group samples \cite{RSB}.  
For example, from the Mulchaey \etal\ \shortcite{Mulchaey96} sample considered
by Bryan \shortcite{Bryan}, the groups with the smallest $R_x$ have among the 
highest \fccool\ values.  On the other hand, the two groups for
which $R_x>0.5$ Mpc have the lowest \fccool.   This emphasizes
the importance of ensuring the X-ray data extend out to at least a sizable fraction
of the virial radius, as we have done when considering the Roussel \etal\ compilation.
We have shown that
there is no discernable trend of \fccool\ with temperature in this
sample, particularly if the baryon fraction is assumed from nucleosynthesis
arguments, rather than relying on the measurement of $M_{\rm hot}$.

\begin{figure}
\begin{center}
\leavevmode \epsfysize=8cm \epsfbox{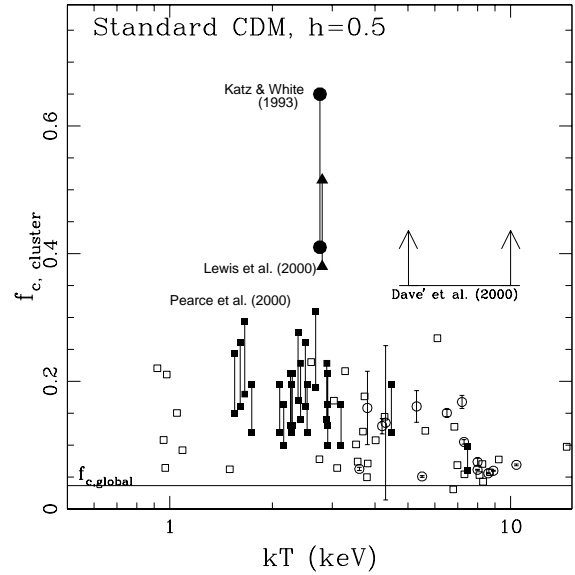}
\end{center}
\caption{The {\it solid} points are the cooled baryon fractions in
several standard CDM simulations, as labelled. The lower point refers to the measured
cooled fraction, and the upper point to which it is joined 
is corrected to include the contribution of unresolved galaxies.  
All of these simulations use $h=0.5$, and cannot be rescaled for
other values of $h$.  The Dav\'e \etal\ (2000) work considers both treecode and 
Eularian grid-based
simulations in several cosmologies; we represent the average cooled fraction of these
simulations as a lower limit, because it does not include cooled gas in
unresolved galaxies.  The data are shown as {\it open symbols}, and
are the same as those in the bottom panel of Figure \ref{fig-clus_obs},
but for $h=0.5$ instead of $h=0.7$.
\label{fig-clus_sims}}
\end{figure}

\section{Numerical Simulations}\label{sec-sims}
Hydrodynamical numerical simulations which attempt to follow the
effects of gas cooling have proved very difficult to
perform across the full range of the gravitational mass
hierarchy. Allowing gas to dissipate its energy via radiative
cooling generates enormous density contrasts and a huge range of
interesting spatial scales within the model, even before the
intrinsically sub-resolution processes of star formation and the
feedback of energy due to supernovae and stellar winds are considered.

The earliest attempt at such a simulation was made by 
Thomas \& Couchman \shortcite{TC92}, though the
model which included radiative cooling was only briefly discussed.
This model had a
very high mass resolution threshold, allowing only the largest
galaxies to form, and only a small fraction of the gas to cool.
This was quickly followed by the work of Katz \& White \shortcite{KW93} who produced
a model that was several years ahead of its time. They demonstrated
that the formation of a Virgo-like cluster, including the
effects of cooling, could be followed by a
computer simulation. However, the model had some problems: in particular,
as the authors make clear, much more gas cools than is observed.
These results are shown as the large, filled circles in Figure \ref{fig-clus_sims}.
The lower point shows the actual fraction of cold baryons in the simulation;
the higher point to which it is connected shows this fraction corrected
by extrapolating the simulated luminosity function to include
the additional cooled gas within unresolved objects, as described
in Katz \& White.  The simulation of Katz \& White was stopped at $z=0.13$,
so the redshift zero fraction will be even higher.

Recently there has been renewed interest in this problem. Suginohara
\& Ostriker \shortcite{SO98} also produced a cluster model, and noted that
additional physical processes (such as star formation and feedback) 
would be required to
prevent the cooling of an excessively large fraction of the gas;
unfortunately, the fraction of cooled baryons was not published in
their paper so we cannot consider their results in Figure \ref{fig-clus_sims}. 
Their work was
followed by that of Pearce \etal \shortcite{Pearce99,Pearce}, Lewis \etal
\shortcite{Lewis-sim} and Dav\'e \etal (2000).  Lewis \etal\ used a
more advanced code to improve upon the earlier simulation of Katz \&
White \shortcite{KW93}, and they also found that around 40\% of the gas within the cluster
cools.  This is shown in Figure \ref{fig-clus_sims} as filled
triangles; the upper triangle includes a correction for galaxies below
the resolution limit, by extrapolating the resultant luminosity function 
assuming a faint end slope $\alpha=-0.96$.  Dav\'e \etal\ used both a
parallel treecode and a Eulerian grid based code in several
cosmologies and, although the cluster-by-cluster fractions are not
quoted, find a universal cooling fraction of around 30\% to 40\% for their high
resolution simulations, in both
cases.  This is shown as a lower limit in Figure \ref{fig-clus_sims}, since
it does not include a correction for galaxies formed in haloes below the
resolution limit.

Simulations like those discussed above which cool too much
gas cannot be expected to model galaxy formation in a  physically correct way.  An alternative,
heuristic approach can be taken to force the simulations to satisfy the constraint on
\fgcool, by limiting the resolution (see \S~\ref{sec-res}).  This is the method
used by Pearce \etal \shortcite{Pearce}, who deliberately set
their mass resolution to cool $\sim$15\% of the gas
within their simulation volume; the cooled fractions for 
the twenty most massive of these
clusters are shown in Figure \ref{fig-clus_sims} as the filled
squares (both before and after correction for sub-resolution
galaxies). By construction, the cooled gas fractions of these simulated haloes
are closer to the observed values, though they are still probably too high.  
The obvious problem with this simulation is that it does not allow the
formation of small galaxies.

\subsection{Resolution}\label{sec-res}
\begin{figure}
\begin{center}
\leavevmode \epsfysize=8cm \epsfbox{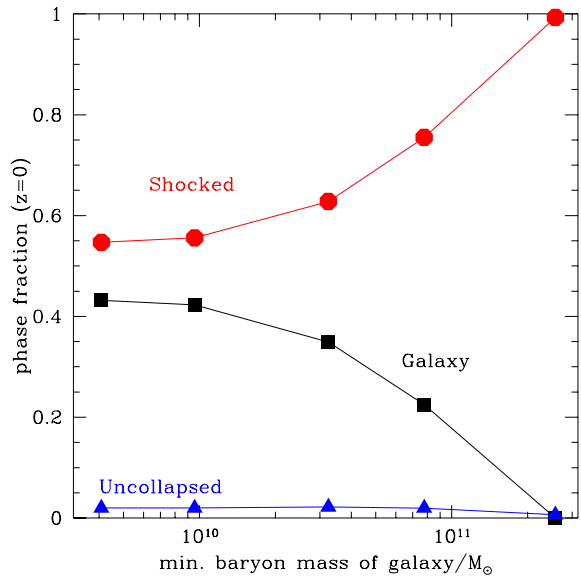}
\end{center}
\caption{Fraction of baryons in three phases (representing shocked gas,
cooled gas [galaxies] and unshocked, diffuse baryons) against the baryonic mass of the smallest resolvable galaxy
for standard CDM simulations with radiative cooling.  The simulations
use a box--size of $10 h^{-1}\,{\rm Mpc}$ and
$\Omega_{\rm b}=0.06$. The two points with the largest mass threshold
are too low, due to artificial heating effects.
Further details are contained in Kay (2000).
\label{fig-coolfrac}}
\end{figure}

It has long been theorised that progessively increasing the resolution
within a simulation leads to a continual increase in the fraction of
material which cools \cite{Cole91,WF91,SO98}, a situation commonly referred to as
the cooling catastrophe. We demonstrate this in
Figure~\ref{fig-coolfrac}, by presenting the results of
standard cold dark matter simulations, at different resolutions 
(the mass of the smallest object that can effectively cool). This figure shows how the fraction of gas
particles in the cold phase increases as resolution is increased.
At the low resolution end, the relation is too steep, due to the artificial
heating problems described by Steinmetz \& White \shortcite{SW97}.  However,
the trends discussed below are qualitatively retained, even when this
problem is overcome \cite{Kay}.
Apart from this effect, the shape of the curve will be approximately independent of
the simulations. 
However, the scale along the abscissa will vary;
that is, the curves may be translated laterally, for example, by
changing $\Omega_b$.  More subtly, in a model
with a positive cosmological constant, structure forms earlier and so
more gas cools at the same threshold mass. This causes the curves
plotted on Figure~\ref{fig-coolfrac} to move to the right by about
a factor of two.  Changes in metallicity can affect the result in the same way,
though even primoridal abundances cannot prevent the cooling 
catastrophe \cite{Dave}.  
In addition, differences in the numerical implementation can
also shift the curve by factors of order two. 


Figure~\ref{fig-coolfrac} makes it clear why some groups find
cooled gas fractions around 40\% whilst others find much lower
fractions. Values around 40\% will be achieved for a wide range of
simulation resolutions, as this is the value obtained when the resolution
mass is less than the characteristic mass of 
the halo mass function. More gas will cool as the resolution
is increased but a large increase is required to see a significant
change, since the slope of the mass function is shallow in this regime.
Any required cold gas fraction
below 40\% can be obtained by choosing the resolution appropriately,
although care must be taken to account for artificial heating \cite{SW97}. 
However, in order to get a good match to the observed
gas fraction within a standard cold dark matter simulation the
resolution limit must be set to $\sim 10^{11}\Msun$ or higher, which
is clearly unphysically high.  This is the situation we referred to
as the ``cooling crisis''; it is a crisis because its resolution will
result in a significant improvement in the utility of simulation results.

\section{Discussion}\label{sec-disc}
We have demonstrated that it is easy to exceed the observed cooled gas
fraction with numerical simulations of sufficient resolution. The
failure to match this constraint reflects the well-known need for
some method of reheating the gas in order to prevent excessive star
formation at high redshift \cite{Larson74,WR78,WF91}.  
Until an effective feedback 
model is implemented in simulations,
it will not be possible to obtain physically illuminating results.

Present attempts to incorporate feedback within cosmological simulations
all suffer from the same drawback, that the physical processes responsible 
for the redistribution of energy occur on scales well below those that can 
presently be resolved. Hence, an accurate treatment of the propagation of
energy from the local ISM to super-galactic scales is currently an 
intractable problem. However, several groups have attempted to include
feedback by injecting the energy into regions at the resolution limit of
the simulation. For example, an approach that has been taken in smoothed
particle hydrodynamics (SPH) 
simulations is simply to add available energy from newly-formed "star"
particles to the thermal energy of the surrounding gas 
\cite{KWH,Lewis-sim,Dave}. 
However, the majority of this gas is cold and dense, and so any excess
energy is rapidly re-radiated. It is plausible that this problem is due
to the failure of the SPH simulations to resolve a multiphase medium, in which 
the feedback would be able to create hot bubbles of diffuse gas, with 
significantly longer cooling times than the cold dense gas from which 
they were spawned. 
Several groups have attempted to "correct" for this by preventing reheated 
gas from radiatively cooling over a finite timescale
(e.g. Gerritsen 1997\nocite{Gerritsen}; Springel 2000\nocite{Springel}; Thacker \& Couchman 2000\nocite{TC00}).
Another method is to supply energy in 
kinetic form (e.g. Navarro \& White 1993\nocite{NW93}; Kay 2000\nocite{Kay}), which is efficient at 
limiting the fraction of cooled gas.  However, the effectiveness of this implementation
may be artificial since the SPH simulations fail to resolve shocks 
that would efficiently thermalize
the reheated material \cite{KWH}. 

An alternative approach to feedback is that 
of  Cen \& Ostriker \shortcite{CO99}, who use a grid-based cosmological
code, and implement analytic rules, similar to those of semi-analytic
models \cite{SP99,Cole2000,KCDW,SPF} to regulate the formation of galaxies
within a single cell.  Unfortunately, the resolution of 
grid-based methods such as this is still too coarse ($\sim 100h^{-1}$ kpc in
the highest resolution simulation of Cen \& Ostriker) to provide useful
results on galaxy scales.  It is not clear that the analytic rules
implemented to model the sub-grid physics in this code are robust to
changes in resolution, as changes in the cell size can result in large
changes in the gas densities, for which the analytic
formalism may not appropriately compensate.
A more promising route may be the use of adaptive-mesh refinement codes
\cite{BN,BV01}, which will allow the implementation of an analytic treatment
of gas cooling within cells of sufficient resolution.  

\begin{figure}
\begin{center}
\epsfig{figure=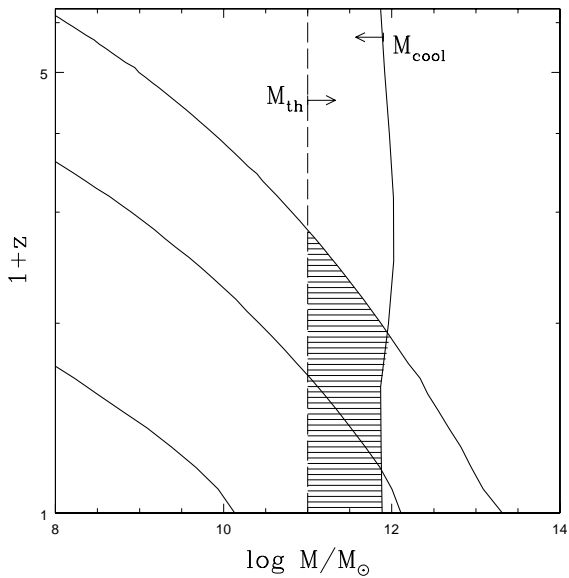,height=8cm,angle=0}
\end{center}
\caption{The region of efficient gas cooling in the Universe,
as a function of halo mass and redshift.  The smooth curved
lines delineate the formation of structure in the Universe.  The
central line represents the median halo mass which virialises at the corresponding
redshift; 50\% of the mass in the Universe collapses between the two lines
which bracket this one.
$M_{\rm cool}$ is the mass threshold below which haloes can cool out to 
their virial radius.  $M_{\rm th}$ is the mass threshold within
the numerical model, generally given by the resolution, below which
cooling cannot occur. The shaded region therefore denotes the region within which
cooling is important.  
\label{fig-scaling}}
\end{figure}

The goal now is to
find a physically motivated feedback model which is able to sufficiently
reduce \fcool.  It is therefore useful to consider, qualitatively, what
is required of such a model.  
The amount of gas which can cool in a given halo cannot be computed 
by considering a halo structure at a single redshift and treating it as if
it had existed unchanged over a Hubble time.  For example, the cooling time
in clusters today is much longer than the Hubble time, and this has been used to
explain why not all of the gas in these environments has cooled \cite{Binney77,Silk77,RO}.
However, this argument does not hold in hierarchical models, where cluster
progenitors at earlier epochs were low mass haloes in which the cooling time was short,
and simulations of hierarchical cluster formation demonstrate that a large fraction of the
baryons cool by the present day.  To treat the problem analytically, it is necessary 
to consider the hierarchical growth of a halo through the plane of Figure \ref{fig-scaling}.
The smooth curved lines represent the evolution of
structure, for a standard CDM model ($\sigma_8=0.8$, $\Gamma=\Omega_\circ h=0.5$); 
the central line shows the median mass halo which virializes as a function
of redshift, and the two others bracket the region in which 50\% of the
mass in the Universe collapses into haloes.   
In a manner similar to that of Rees \& Ostriker \shortcite{RO},
we can define an efficient cooling mass, $M_{\rm cool}$,
as the mass which can cool all the gas out to its virial radius, within
a Hubble time at redshift $z$. 
The cooling rate, which depends on halo density and temperature,
is computed using the models of Raymond \etal \shortcite{RCS}, with
one-third solar metallicity.  
This cooling mass, for the standard CDM model described above, is shown
in Figure \ref{fig-scaling}, and is given by $M_{\rm cool} \sim 10^{12}$\Msun, 
almost completely independent of
redshift for $z<10$.  This non-trivial result is a consequence of the fact that,
when cooling is dominated by line emission, the increase in the cooling rate with increasing redshift
is closely balanced by the decrease in the Hubble time.
This forms an approximate upper mass limit
to efficient cooling;  haloes with masses smaller than this
limit can cool all of their gas, out to the virial radius, while to the right
of this line, only a rapidly decreasing fraction of the gas is able to cool.

At low redshifts, most of the haloes which are collapsing are more
massive than $M_{\rm cool}$; therefore,
cooling plays a minor role today.  By $z\approx1$, however, the characteristic
mass crosses the $M_{\rm cool}$ threshold, and cooling becomes a dominant process,
leading to the cooling catastrophe.  In simulations, the resolution threshold
imposes a mass limit, $M_{\rm th}$, which prevents smaller mass haloes from cooling.
Efficient cooling is therefore restricted to the shaded region in Figure \ref{fig-scaling}.
As $M_{\rm th}$ is lowered, more and more mass cools, and will quickly 
exceed the tight observational constraints.  If the threshold is high enough
to prevent efficient cooling at high redshift, no low mass galaxies
are formed. 

The final fraction of cooled gas then depends on the trajectory a particular
halo takes through this plane, as it grows.  This growth will occur
via both smooth accretion, and discrete jumps due to mergers.
The resolution threshold imprints a scale on the
accumulation of cooled baryons, as the final fraction \fcool\ in a given
halo will depend on its trajectory only after it becomes sufficiently
resolved.  This could have an effect on the dependence of \fcool\ on halo
mass in simulations (but see Kay 2000).

It is clear that any analytic feedback model which
behaves in a manner similar to a sharp, fixed mass threshold will fail to
simultaneously match the observed \fgcool\ and still produce a reasonable galaxy luminosity function.
A physical feedback scheme, however, does not provide a hard mass threshold,
but gradually alters the cooling efficiency of haloes as a function of mass and redshift;
thus the shaded region in Figure \ref{fig-scaling} becomes a ``fuzzy'' region
without a well defined lower boundary.  There are many models for such feedback, including
supernovae or AGN reheating \cite{Larson74,WR78,LS91,Cole91,Wu,Bower01,MLF,E2000},
ionizing radiation from the ultraviolet background \cite{UV,WHK}, and an
entropy threshold established at high redshift \cite{K91,BVM,Ponman,entropy}.
A very successful implementation of a feedback scaling law is that
used by semi-analytic models of galaxy formation \cite{Cole2000,SP99,KCDW,SPF},
in which the mass of reheated gas is assumed to scale in proportion
with the halo circular velocity raised to some power $\alpha$.  Though this
scaling is constructed to allow the models to match the observed luminosity
function, it also succeeds in limiting \fcool\ to more reasonable values
(though generally still high compared with the constraints described here).
Furthermore, there may be some theoretical motivation for this scaling relation
from supernova-driven wind models \cite{E2000}.
The next step in this work is to combine these models with realistic halo merger trees 
to determine, analytically, how they affect the cooling rate as a function of halo
mass and redshift, for implementation into numerical simulations. 

\section{Conclusions}
In this paper, we have presented an up-to-date review of the cooling
catastrophe.  Although the issues we have discussed have been known
for a long time, recent advances in observations and simulation techniques
warrant a reevaluation of the situation, which can be summarized in
two key points:
\begin{itemize}
\item The global fraction of baryons in the form of stars and cold gas
 is now well constrained observationally from the $K-$band luminosity function
to be \fgcool$\approx 0.073h$. 
Although there is some uncertainty
about how this may vary with halo mass, there is no convincing evidence for
a significant difference from the global value.  
\item The amount of gas which cools in simulations is directly
related to the resolution, unless a working reheating scheme is
implemented.  No current simulation is able to produce both
(1) the correct \fgcool\ and (2) a population of low mass galaxies,
without implementing analytic models of the sub-grid physics
on uncomfortably large scales.
\end{itemize}

The use of simulations to aid our understanding of galaxy properties
and of cluster scaling laws therefore awaits the implementation of a
sub-grid, physical feedback scheme which allows the models to satisfy observational
constraints on \fgcool\ in a resolution-independent way.  

\section*{Acknowledgements}
MLB acknowledges support from a PPARC rolling grant for extragalactic astronomy
and cosmology at Durham.  We thank Simon White, Eric Bell, Greg Bryan and Cedric Lacey for several suggestions
and corrections which greatly improved this paper.  Finally, we thank the
referee for providing a thorough report which allowed us to improve the
focus and clarity of this work.
\bibliographystyle{astron_mlb}
\bibliography{ms}

\end{document}